\pdfoutput=1
\RequirePackage{ifpdf}
\ifpdf 
\documentclass[pdftex]{sigma}
\else
\documentclass{sigma}
\fi

\usepackage{mathrsfs}

\newcommand{\Tr}{\mathop{\rm Tr}\nolimits}
\newcommand{\BE}{\mathop{\rm BE}\nolimits}
\newcommand{\diag}{\mathop{\rm diag}}
\newcommand{\gen}{\mathop{\rm gen}}

\begin{document}

\allowdisplaybreaks

\renewcommand{\PaperNumber}{072}

\FirstPageHeading

\ShortArticleName{XXX Model with General Boundaries: Algebraic Bethe Ansatz}

\ArticleName{Heisenberg XXX Model with General Boundaries:\\
Eigenvectors from Algebraic Bethe Ansatz}

\Author{Samuel BELLIARD~$^{\dag\ddag}$ and Nicolas CRAMP\'E~$^{\dag\ddag}$}

\AuthorNameForHeading{S.~Belliard and N.~Cramp\'e}

\Address{$^\dag$~Laboratoire Charles Coulomb L2C, UMR 5221, CNRS, F-34095 Montpellier, France}

\Address{$^\ddag$~Laboratoire Charles Coulomb L2C, UMR 5221,\\
\hphantom{$^\ddag$}~Universit\'e Montpellier 2, F-34095 Montpellier, France}
\EmailD{\href{mailto:samuel.belliard@univ-montp2.fr}{samuel.belliard@univ-montp2.fr},
\href{mailto:nicolas.crampe@univ-montp2.fr}{nicolas.crampe@univ-montp2.fr}}

\ArticleDates{Received September 29, 2013, in f\/inal form November 19, 2013; Published online November 22, 2013}

\Abstract{We propose a~generalization of the algebraic Bethe ansatz to obtain the eigenvectors of the
Heisenberg spin chain with general boundaries associated to the eigenvalues and the Bethe equations found
recently by Cao et al.
The ansatz takes the usual form of a~product of operators acting on a~particular vector except that the
number of operators is equal to the length of the chain.
We prove this result for the chains with small length.
We obtain also an of\/f-shell equation (i.e.\ satisf\/ied without the Bethe equations) formally similar to
the ones obtained in the periodic case or with diagonal boundaries.}

\Keywords{algebraic Bethe ansatz; integrable spin chain; boundary conditions}

\Classification{82B23; 81R12}

\section{Introduction}

The Heisenberg XXX spin chain of length $N$ with open boundaries is described by the following Hamiltonian
\begin{gather}\label{eq:H}
H=\frac{1}{q}\left(\sigma^{z}_{1}+\xi^+\sigma^{+}_{1}+\xi^-\sigma^{-}_{1}\right)+\sum_{n=1}^{N-1}
\vec\sigma_{n}\cdot\vec\sigma_{n+1}+\frac{1}{p}\left(\sigma^{z}_{N}+\eta^+\sigma^{+}_{N}+\eta^-\sigma^{-}_{N}\right),
\end{gather}
where $\sigma^i$ stands for the standard Pauli matrix and $q$, $\xi^+$, $\xi^-$, $p$, $\eta^+$, $\eta^-$
are arbitrary parameters characterizing the boundaries.
This model is one of the simplest open integrable models and f\/inds applications in a~wide range of
domains such that: condensed matter physics, high energy physics, mathematical physics, \dots.

The study of the spectral problem associated to this model has started in 1983 with the Gaudin's
book~\cite{gaudinB} where the coordinate Bethe ansatz (CBA), introduced by Bethe~\cite{Bet}, was applied
for diagonal boundaries (i.e.\ $\xi^\pm=\eta^\pm=0$).
In 1988, the generalization of the algebraic Bethe ansatz (ABA), developed by Faddeev's school~\cite{SFT},
to deal with open boundaries was introduced by Sklyanin~\cite{sklyanin} and he recovered Gaudin's results.
The f\/irst step in both constructions is the identif\/ication of one particular eigenvector, usually
called the pseudo-vacuum.

More importantly, the ABA opens the way to the algebraic study of XXX open integrable model from the
ref\/lection equation~\cite{chered,sklyanin}
\begin{gather}\label{eq:re}
R_{12}(u-v)K_{1}(u)R_{12}(u+v)K_{2}(v)=K_{2}(v)R_{12}(u+v)K_{1}(u)R_{12}(u-v),
\end{gather}
where $R(u)$ is the rational $R$-matrix solution of Yang--Baxter equation~\cite{baxter3, yang} and $K(u)$ is
the open monodromy matrix.
This leads one, using the scalar solution (i.e.\ acting trivially in the spin spaces) of the ref\/lection
equation found in~\cite{Kmat, GZ} and the open monodromy matrix, to the construction of the transfer matrix
$t(u)$ which is the generating function of the Hamiltonian and the conserved quantities of the model.
This allows one to prove the integrability of the XXX model with generic boundaries.

The integrability for generic boundaries being proven, we expect that the spectral problem can be solved
exactly.
However, in this case, there is no obvious pseudo-vacuum for the Hamiltonian or for the transfer matrix.
This is due to the fact that the boundaries break the ${\rm U}(1)$-symmetry (i.e.\ the total spin is not anymore
conserved).
This problem is quite general for integrable models without ${\rm U}(1)$-symmetry and led numerous researchers to
believe that usual methods cannot work in this case and to develop dif\/ferent approaches to study their
spectrum over the last 30 years.
Basically, one can consider two families of approaches:
\begin{itemize}\itemsep=0pt
\item to modify the various Bethe ansatz and parametrize the spectrum by a~set of Bethe roots
that must satisfy Bethe equations: CBA~\cite{CR, CRS1,CRS2}, ABA~\cite{BCR13,CLSW,MMR,pimenta,YanZ07}, $T$-$Q$
equations~\cite{CYSW2,CaoYSW13-4,MurN05, nepo} and functional Bethe ansatz~\cite{FraGSW11,FSW, Gal08};
\item
to obtain the spectrum without use of the Bethe equations: separation of variables~\cite{FKN, niccoli2},
$q$-Onsager approach~\cite{BB,BK} and matrix ansatz~\cite{GLMV,Laz}.
\end{itemize}

Recently, progress on the generalization of $T$-$Q$ relation formalism has been performed by Cao, Yang, Shi and
Wang~\cite{CYSW} where the eigenvalues and the Bethe equations of Hamiltonian~\eqref{eq:H} are obtained.
The main feature consists in adding a~new term in the eigenvalues and the Bethe equations in comparison to
the usual ones obtained for cyclic or diagonal boundaries (see also~\cite{frahm} for this idea).
Then, this result has been simplif\/ied by Nepomechie~\cite{Nep} that shows numerical evidences for the
completeness of the spectrum and that the suf\/f\/icient and necessary number of Bethe roots is~$N$, the
length of the chain\footnote{In~\cite{CYSW}, two other supplementary sets of Bethe roots with dimension
$[N/2]$ are also introduced.}.

Another recent result~\cite{BCR13} is the generalization of the ABA for the case with upper triangular
boundary (i.e.\ $\eta^-=\xi^-=0$) which is already a~case without ${\rm U}(1)$-symmetry.
The Bethe vectors introduced in this case are superpositions of the ones, with dif\/ferent total spin, used
in the case with diagonal boundary.
This idea to consider a~superposition of the vectors of the diagonal case was used f\/irst in the context
of the CBA~\cite{CR, CRS1,CRS2} then in the context of the ABA for the XXZ spin chain~\cite{pimenta}.

These two results permit us to propose a~generalization of the ABA to get the eigenvectors (see
equation~\eqref{eq:eigen}) of the Hamiltonian~\eqref{eq:H} for generic boundaries.
The eigenvectors constructed by ABA is usually a~crucial step forward to get the correlation functions of
the model (see~\cite{BPRS12, KitMT99} for periodic case or~\cite{KitKMNST07,KitKMNST07-II} for diagonal
case).
It is also the method which can be the most simply generalized to obtain the eigenvectors for the $s$-spin
chain.
In addition, we show that the action of the transfer matrix on the Bethe vectors is given by a~``wanted''
term and an ``unwanted'' term (see equation~\eqref{eq:vpBE}) as in the diagonal case.

The paper is organized as follows.
In Section~\ref{sec2}, we f\/ix the notations and recall the eigenvalues and the Bethe equations obtained
in~\cite{CYSW}.
Then, in Section~\ref{sec3}, we give our conjecture for the Bethe vectors of the model and discuss the new technical
problems for a~general proof of our result.
Section~\ref{sec:smallL} presents the proof for chain with small length $N=1,2,3$ to support our conjecture.
Finally, in Section~\ref{sec5}, we discuss some perspectives and possible generalizations of our result to other
models such that the XXZ spin chain.

\section{Transfer matrix and eigenvalues}\label{sec2}

In this section, we recall the construction of the transfer matrix used to prove the integrability of the
Hamiltonian and f\/ix the notations of the main objects used in this paper.
Then, we present the eigenvalues and the Bethe equations obtained in~\cite{CYSW,Nep}.

The invariance of the Hamiltonian~\eqref{eq:H} by conjugation allows one to consider $\eta^\pm=0$ without
loss of generality.
Let us emphasize that we study a~slightly more general Hamiltonian than the one studied in~\cite{CYSW,Nep}
(recovered by setting $\xi^+=\xi^-=\xi$) such that the triangular case can be obtained easily as limit.

The main object in the context of quantum integrable models is the $R$-matrix, solution of the Yang--Baxter
equation~\cite{baxter3, yang}.
In the case of XXX spin chain, the associated $R$-matrix is called the rational $R$-matrix and is given by
\begin{gather*}
R(u)=u+{\cal P},
\end{gather*}
where ${\cal P}$ is the permutation matrix.
This $R$-matrix is invariant, i.e.\ $[R(u),Q\otimes Q]=0$ for any invertible 2 by 2 matrix $Q$.
This invariance allows one to show that $Q^{-1}K(u) Q$ is a~solution of the ref\/lection
equation~\eqref{eq:re} if $K(u)$ is also a~solution (one conjugates the ref\/lection equation~\eqref{eq:re}
for $K$ by $Q_1Q_2$ and moves the $Q$ using the invariance of $R$).

The $R$-matrix allows one to construct the following transfer matrix~\cite{sklyanin}
\begin{gather}
t(u)=\Tr_{0}\big(K^{+}_{0}(u)K_{0}(u)\big),
\qquad
K_{0}(u)=T_{0}(u)K^{-}_{0}(u)\hat T_{0}(u),
\label{transfer}
\end{gather}
where the open monodromy matrix is built by Sklyanin dressing procedure with bulk monodromy matrices
\begin{gather*}
T_{0}(u)=R_{01}(u-\theta_{1})\cdots R_{0N}(u-\theta_{N}),
\qquad
\hat T_{0}(u)=R_{0N}(u+\theta_{N})\cdots R_{01}(u+\theta_{1}),
\end{gather*}
and the $K^-$-matrices, scalar solutions of the ref\/lection equation~\eqref{eq:re}\footnote{To be precise,
$K^+$ is solution of the dual ref\/lection equation.}, are~\cite{Kmat, GZ}
\begin{gather*}
K^{-}(u)=\left(
\begin{matrix}
p+u&0
\\
0&p-u
\end{matrix}
\right),
\qquad
K^{+}(u)=\left(
\begin{matrix}
q+u+1&\xi^+(u+1)
\\
\xi^-(u+1)&q-u-1
\end{matrix}
\right).
\end{gather*}
The parameters $\theta_{1}, \ldots, \theta_{N}$ are free parameters called inhomogeneity parameters.
Two transfer matrices at dif\/ferent values of their spectral parameter commute (i.e.\ $[t(u),t(v)]=0$).
Therefore, the expansion of the transfer matrix $t(u)$ w.r.t.\ the spectral parameter $u$ provides a~set of
commuting operators among which is the Hamiltonian~\eqref{eq:H}.
This proves the integrability of this Hamiltonian.

In addition to that, the above procedure allows one to construct the main operators necessary in the
context of the algebraic Bethe ansatz as well as to get the exchange relations between them.
Indeed, these operators are the entries of the open monodromy matrix, dressed solution of the ref\/lection
equation~\eqref{eq:re},
\begin{gather}
K_0(u)=\left(
\begin{matrix}
\mathscr{A}(u)&\mathscr{B}(u)
\\
\mathscr{C}(u)&\mathscr{D}(u)+\frac{1}{2u+1}\mathscr{A}(u)
\end{matrix}
\right).
\label{eqB}
\end{gather}
Then, the transfer matrix~\eqref{transfer} may be rewritten\footnote{Let us emphasize that we choose
a~dif\/ferent normalization for our operator $\mathscr{D}$ in comparison to~\cite{CYSW} (where it is called
$\bar D$).
Precisely, on get ${\bar D}(u)=(2u+1)\mathscr{D}(u)$.
It explains the discrepancies between some of the relations given here and the corresponding ones given
in~\cite{CYSW}.}
\begin{gather}\label{trane}
t(u)=\frac{2(u+q)(u+1)}{2u+1}\mathscr{A}(u)+(u+1)\big(\xi^-\mathscr{B}(u)+\xi^+\mathscr{C}
(u)\big)+(q-u-1)\mathscr{D}(u).
\end{gather}
Since the monodromy matrix $K(u)$ given by~\eqref{eqB} satisf\/ies the ref\/lection
equation~\cite{sklyanin}, the exchange relations between $\mathscr{A}(u)$, $\mathscr{B}(u)$,
$\mathscr{C}(u)$ and $\mathscr{D}(u)$ can be computed.
We gather the ones used in this paper in Appendix~\ref{App:cr}.
In addition to that, $K^-$ being diagonal, the actions of these operators on the pseudo-vacuum
$|\Omega\rangle$ (i.e.\ the vector with all spins up) are the usual ones:
\begin{alignat}{3}
& \mathscr{A}(u)|\Omega\rangle=\Lambda_1(u)|\Omega\rangle,
\qquad &&
\mathscr{D}(u)|\Omega\rangle=\Lambda_2(u)|\Omega\rangle, &
\nonumber
\\
& \mathscr{C}(u)|\Omega\rangle=0,
\qquad &&
\mathscr{B}(u)\mathscr{B}(\lambda_1)\cdots\mathscr{B}(\lambda_N)|\Omega\rangle=0, &
\label{eq:act}
\end{alignat}
where
\begin{gather*}
\Lambda_1(u)=(u+p)\prod_{j=1}^{N}\big((u+1)^2-\theta_j^2\big)
\qquad
\text{and}
\qquad
\Lambda_2(u)=\frac{2u}{2u+1}(p-u-1)\prod_{j=1}^{N}\big(u^2-\theta_j^2\big).
\end{gather*}

Recently, in~\cite{CYSW}, using inhomogeneous $T$-$Q$ equations, the eigenvalues of the transfer
mat\-rix~\eqref{transfer} have been computed.
Then, in~\cite{Nep}, supported by numerical evidences for chains of small length, it is shown that only
sub-cases are suf\/f\/icient to get the complete spectrum.
In our notations, these eigenvalues can be written as follows
\begin{gather}
\Lambda(u)=\bar{\alpha}(u)\Lambda_1(u)\prod_{j=1}^N f(u,\lambda_j)+\bar{\delta}(u)\Lambda_2(u)\prod_{j=1}
^N h(u,\lambda_j)
\nonumber
\\
\phantom{\Lambda(u)=}
+\rho\frac{(u+1)(2u+1)}{(u+p)(p-u-1)}\Lambda_1(u)\Lambda_2(u)\prod_{j=1}^N\frac{1}
{(u-\lambda_j)(u+\lambda_j+1)},\label{eq:L}
\end{gather}
where $f$ and $h$ are def\/ined in~\eqref{eq:fg}, $\rho=1-\sqrt{1+\xi^+\xi^-}$ and
\begin{gather}\label{eq:ad}
\bar{\alpha}(u)=\frac{2(u+1)}{2u+1}((1-\rho)u+q),
\qquad
\bar{\delta}(u)=q-(u+1)(1-\rho).
\end{gather}
The set of parameters $\{\lambda_k\}$ are called Bethe roots and must satisfy the Bethe equations to assure
that the residues of $\Lambda(u)$ at $u=\lambda_k$ vanish.
One def\/ines\footnote{Dif\/ferent normalizations are possible but we chose the most convenient for the
following computations.}
\begin{gather*}
\BE_k=-2\lambda_k\bar{\alpha}(\lambda_k)\Lambda_1(\lambda_k)\prod_{\genfrac{}{}{0pt}{}{j=1}{j\neq k}
}^Nf(\lambda_k,\lambda_j)+2(\lambda_k+1)\bar{\delta}(\lambda_k)\Lambda_2(\lambda_k)\prod_{\genfrac{}{}{0pt}
{}{j=1}{j\neq k}}^N h(\lambda_k,\lambda_j)
\\
\phantom{\BE_k=}
{}+\rho\frac{(\lambda_k+1)(2\lambda_k+1)}{(\lambda_k+p)(p-\lambda_k-1)}
\Lambda_1(\lambda_k)\Lambda_2(\lambda_k)\prod_{\genfrac{}{}{0pt}{}{j=1}{j\neq k}}^N\frac{1}
{(\lambda_k-\lambda_j)(\lambda_k+\lambda_j+1)}.
\end{gather*}
Then, the Bethe equations are
\begin{gather}\label{eq:BEK}
\BE_k=0
\qquad
\text{for}
\quad
k=1,2,\dots,N.
\end{gather}
For diagonal left boundary (i.e.\ $\xi^+=\xi^-=\rho=0$), we recover the eigenvalues, denoted
$\Lambda^{\diag}(u)$, and the Bethe equations, denoted $\BE_k^{\diag}$, computed in~\cite{sklyanin}.
This allows us to rewrite the eigenvalues and the Bethe equations in a~suitable form for the following
computations:
\begin{gather}\label{eq:fr}
\Lambda(u)=\Lambda^{\diag}(u)+\rho\Lambda^{\gen}(u)
\qquad
\text{and}
\qquad
\BE_k=\BE_k^{\diag}+\rho\BE_k^{\gen}.
\end{gather}

\section{Algebraic Bethe ansatz}\label{sec3}

In this section, we give the main results of this letter: the eigenvectors of the transfer
matrix~\eqref{transfer} with the eigenvalues~\eqref{eq:L} using the algebraic Bethe ansatz.

Before that, let recall the usual construction of the Bethe vectors for diagonal boundaries (i.e.\
$\xi^+=\xi^-=0$)~\cite{sklyanin}.
In this case, there is a~${\rm U}(1)$-symmetry (the total spin is conserved), therefore the Hamiltonian can be
decomposed into a~direct sum of dif\/ferent sectors with a~given spin.
In the sector with spin $N/2-M$, the diagonal Bethe vectors read
\begin{gather}\label{eq:fd}
|\Phi^M(\lambda_1,\lambda_2,\dots,\lambda_M)\rangle^{\diag}=\mathscr{B}(\lambda_1)\mathscr{B}
(\lambda_2)\cdots\mathscr{B}(\lambda_M)|\Omega\rangle.
\end{gather}
They are the eigenvectors of the transfer matrix if $\{\lambda_1,\lambda_2,\dots,\lambda_M\}$ satisfy the
diagonal Bethe equations.
The proof uses the exchange relations of the operators $\mathscr{A}$ and $\mathscr{D}$ with $\mathscr{B}$
as well as their action on the pseudo-vacuum in order to compute the action of the transfer matrix on the
Bethe vectors.
Then, one gets the following result:
\begin{gather}
t^{\diag}(u)|\Phi^M(\lambda_1,\dots,\lambda_M)\rangle^{\diag}=\Lambda^{\diag}
(u)|\Phi^M(\lambda_1,\dots,\lambda_M)\rangle^{\diag}
\nonumber
\\
\qquad
{}+\sum_{k=1}^M F(u,\lambda_k)\BE ^{\diag}_k|\Phi^M(\lambda_1,\dots,\lambda_{k-1},u,\lambda_{k+1}
,\dots,\lambda_M)\rangle^{\diag},\label{eq:offd}
\end{gather}
where
\begin{gather}\label{eq:F}
F(u,\lambda)=\frac{u+1}{(u+\lambda+1)(\lambda-u)(\lambda+1)}.
\end{gather}
Let us emphasize that relation~\eqref{eq:offd} is of\/f-shell, i.e.\ the diagonal Bethe equations are not
necessarily satisf\/ied $\BE ^{\diag}_k\neq 0$.
Then, asking that the $\lambda$ satisfy them, we see that the Bethe vectors become eigenvectors.

For $\xi^+\neq 0$ or $\xi^-\neq 0$, the usual ansatz~\eqref{eq:fd} cannot work since the ${\rm U}(1)$-symmetry is
broken and the total spin is not conserved.
In spite of this problem, we believed that a~generalization of the usual algebraic Bethe ansatz is still
possible to obtain the eigenvectors.
Before giving our proposal (see equation~\eqref{eq:eigen}), we recall the dif\/ferent results which lead us
to it:
\begin{itemize}\itemsep=0pt
\item[$(i)$] As the total spin is not conserved, the appropriate ansatz must be a~linear
combination of Bethe vectors~\eqref{eq:fd} for dif\/ferent values of $M$.
At this point, the problem consists in computing the coef\/f\/icients of this linear combination, called
transmission coef\/f\/icients since it corresponds to a~pseudo-particle transmits to a~reservoir.
Nothing, a~priori, guarantees that the transmission coef\/f\/icients take a~simple form.
However, this problem appears already for triangular boundaries ($\xi^+=0$ but $\xi^-\neq 0$) and,
in~\cite{CR}, in the context of the coordinate Bethe ansatz or, in~\cite{BCR13}, in the context of the
algebraic Bethe ansatz, these coef\/f\/icients have been successfully computed and are simple.
\item[$(ii)$]
The eigenvalues and the Bethe equations obtained in~\cite{CYSW} are not so dif\/ferent from the ones for
the diagonal boundaries: they are the sum of the diagonal ones and an additional term which is a~function
of the of\/f-diagonal terms (see equations~\eqref{eq:fr}).
This indicates that, maybe, a~slight generalization of the algebraic Bethe ansatz is enough to obtain the
eigenvectors.
\item[$(iii)$] As explained previously, one set of $N$ Bethe roots is enough to obtain the
complete set of eigenvalues~\cite{Nep}.
This suggests that the ansatz must have a~maximum of $N$ operators.
Moreover, our experience on the triangular case~\cite{BCR13} taught us that the presence of the operator
$\mathscr{C}$ in the transfer matrix requires the superposition of diagonal Bethe vectors belonging to
sectors of spins from $N/2$ to $N/2-M$ with $M\leq N$.
In the general case, the addition of the operator $\mathscr{B}$ in the transfer matrix indicates we need
a~superposition of diagonal Bethe vectors from each sector.
\end{itemize}

These dif\/ferent points and the investigation of the spin chains with small length lead us to the
following ansatz for the Bethe vectors
\begin{gather}\label{eq:eigen}
|\Phi^N(\lambda_1,\lambda_2,\dots,\lambda_N)\rangle=\overline{\mathscr{B}}(\lambda_1)\overline{\mathscr{B}}
(\lambda_2)\cdots\overline{\mathscr{B}}(\lambda_N)|\Omega\rangle,
\end{gather}
where
\begin{gather}\label{eq:defbar}
\overline{\mathscr{B}}(\lambda)=\mathscr{B}(\lambda)+\frac{\rho}{\xi^-}\left(\frac{2\lambda}{2\lambda+1}
\mathscr{A}(\lambda)-\mathscr{D}(\lambda)\right)-\left(\frac{\rho}{\xi^-}\right)^2\mathscr{C}(\lambda).
\end{gather}
We claim that the Bethe vectors~\eqref{eq:eigen} are the eigenvectors of the transfer matrix with the
eigenvalues $\Lambda(u)$ (given by~\eqref{eq:L}) if $\lambda_1,\lambda_2,\dots,\lambda_N$ satisfy the Bethe
equations~\eqref{eq:BEK}.
We do not have yet a~complete proof of this result but it is supported by the study of the chains of length
1, 2 and 3 where the result holds.
We give details of the proof for $N=1,2,3$ in Section~\ref{sec:smallL}.
As for the diagonal boundaries, the Bethe vectors satisfy the following of\/f-shell equation
\begin{gather}
t(u)|\Phi^N(\lambda_1,\dots,\lambda_N)\rangle=\Lambda(u)|\Phi^N(\lambda_1,\dots,\lambda_N)\rangle
\nonumber
\\
\qquad
{}+\sum_{k=1}^N F(u,\lambda_k){\BE }_k|\Phi^N(\lambda_1,\dots,\lambda_{k-1},u,\lambda_{k+1},
\dots,\lambda_N)\rangle,\label{eq:vpBE}
\end{gather}
where $F$ is given by~\eqref{eq:F}.

Let us emphasize that, in this framework, it is straightforward to take the homogeneous limit (i.e.\
$\theta_i=0$) contrary to the separation of variable method.

At this point, we would like to make dif\/ferent remarks on this result, explain why the Bethe
vector~\eqref{eq:eigen} is a~reasonable candidate to be an eigenvector and raise technical problems.

At f\/irst sight, expression~\eqref{eq:defbar} for $\overline{\mathscr{B}}(u)$ seems unnatural.
Nevertheless, we show here that there exists a~simple way to obtain it.
Firstly, let us diagonalize the left $K$-matrix
\begin{gather*}
Q^{-1}K^+(u)Q=D^+(u)=\left(
\begin{matrix}
q+(u+1)(1-\rho)&0
\\
0&q-(u+1)(1-\rho)
\end{matrix}
\right),\qquad
Q=\left(
\begin{matrix}
\xi^+&\rho
\\
-\rho&\xi^-
\end{matrix}
\right).
\end{gather*}
Secondly, thanks to this result and to the fact that the trace is cyclic, the transfer matrix may be
rewritten as follows
\begin{gather*}
t(u)=\frac{(\xi^-)^2}{\xi^+\xi^-+\rho^2}\Tr_0\big(D_0^+(u)\overline{K}_0(u)\big),
\qquad
\overline{K}_0(u)=\frac{\xi^+\xi^-+\rho^2}{(\xi^-)^2}Q_0^{-1}K_0(u)Q_0.
\end{gather*}
Finally, computing the right upper entry of the matrix $Q_0^{-1} K_0(u) Q_0$, we recover, up to a~factor,
the r.h.s.\
of~\eqref{eq:defbar}.
Let us also focus on the fact that the entries of $D^+(u)$ are very similar to the ones appearing in the
eigenvalue via the functions $\bar \alpha$ and $\bar \delta$ (see equation~\eqref{eq:ad}).
For completeness we gives the full set of new generators
\begin{gather*}
\overline{K}(u)=\left(
\begin{matrix}
\overline{\mathscr{A}}(u)&\overline{\mathscr{B}}(u)
\\
\overline{\mathscr{C}}(u)&\overline{\mathscr{D}}(u)+\frac{1}{2u+1}\overline{\mathscr{A}}(u)
\end{matrix}
\right)
\end{gather*}
with
\begin{gather*}
\overline{\mathscr{A}}(u)=\left(\frac{\xi^+}{\xi^-}+\frac{1}{2u+1}\left(\frac{\rho}{\xi^-}
\right)^2\right)\mathscr{A}(u)-\frac{\rho}{\xi^-}\left(\mathscr{B}(u)+\frac{\xi^+}{\xi^-}\mathscr{C}
(u)\right)+\left(\frac{\rho}{\xi^-}\right)^2\mathscr{D}(u),
\\
\overline{\mathscr{D}}(u)=\left(\frac{\xi^+}{\xi^-}-\frac{1}{2u+1}\left(\frac{\rho}{\xi^-}\right)^2\right)\mathscr{D}(u)
+\frac{2(u+1)}{2u+1}\frac{\rho}{\xi^-}\left(\mathscr{B}(u)+\frac{\xi^+}{\xi^-}\mathscr{C}(u)\right)
\\
\phantom{\overline{\mathscr{D}}(u)=}
{}+\frac{4u(u+1)}{(2u+1)^2}\left(\frac{\rho}{\xi^-}\right)^2\mathscr{A}(u),
\\
\overline{\mathscr{C}}(u)=\left(\frac{\xi^+}{\xi^-}\right)^2\mathscr{C}(u)+\frac{\xi^+}{\xi^-}
\left(\frac{2u}{2u+1}\mathscr{A}(u)-\mathscr{B}(u)\right)-\left(\frac{\rho}{\xi^-}\right)^2\mathscr{B}(u).
\end{gather*}
One can remark that
\begin{gather*}
\overline{\mathscr{A}}(u)|\Omega\rangle=\frac{\xi^+\xi^-+\rho^2}{(\xi^-)^2}
\Lambda_1(u)|\Omega\rangle-\frac{\rho}{\xi^-}\overline{\mathscr{B}}(u)|\Omega\rangle,
\\
\overline{\mathscr{D}}(u)|\Omega\rangle=\frac{\xi^+\xi^-+\rho^2}{(\xi^-)^2}
\Lambda_2(u)|\Omega\rangle+\frac{2(u+1)}{2u+1}\frac{\rho}{\xi^-}\overline{\mathscr{B}}(u)|\Omega\rangle
\end{gather*}
and
\begin{gather*}
t(u)=\frac{(\xi^-)^2}{\xi^+\xi^-+\rho^2}\left(\bar\alpha(u)\overline{\mathscr{A}}
(u)+\bar\delta(u)\overline{\mathscr{D}}(u)\right).
\end{gather*}

One knows also that $\overline{K}(u)=Q^{-1} K(u) Q$ satisf\/ies the ref\/lection equation.
Therefore, its entries satisfy the same relations as the entries of $K(u)$ and in particular one gets
\begin{gather*}
\big[\overline{\mathscr{B}}(u),\overline{\mathscr{B}}(v)\big]=0.
\end{gather*}
This allows us to conclude that the Bethe vector $|\Phi^N(\lambda_1,\lambda_2,\dots,\lambda_N)\rangle$ is
completely symmetric on its parameters as in the case with diagonal boundary.

Using exchange relations of Appendix~\ref{App:cr}, the Bethe vectors may be rewritten as follows
\begin{gather}\label{eq:f2}
|\Phi^N(\lambda_1,\dots,\lambda_N)\rangle=\sum_{m=0}^N
\sum_{1\leq i_1<\dots<i_m\leq N}
W_{i_1,\dots,i_m}(\lambda_{1},\dots,\lambda_{N})
\mathscr{B}(\lambda_{i_1})\cdots\mathscr{B}(\lambda_{i_m})|\Omega\rangle,
\end{gather}
where the transmission coef\/f\/icients $W$ can be computed explicitly.
Therefore, we see that our proposal contains vectors with all possible values for the total spin.
This feature was expected as explained above (see remark $(iii)$).

The computation of the explicit form of functions $W$ may be complicated.
Nevertheless, using the recurrence relations introduced to f\/ind the spectrum (see, e.g.,
\cite{CYSW2, Gal08}), it may possible to f\/ind also some recurrence relations between these functions which
would be helpful to compute them.

The transmission coef\/f\/icients $W$ contain some $\Lambda_1(\lambda_j)$ or $\Lambda_2(\lambda_j)$.
Therefore, using the Bethe equations, we can rewrite them dif\/ferently.
Depending on the computations one wants to perform, the most suitable form must be chosen to simplify them.
Let us remark that the of\/f-shell equation~\eqref{eq:vpBE} seems to be valid only for the form obtained by
the direct development of~\eqref{eq:eigen}.

Another point which validates the conjecture is that we recover as a~limit the case with triangular
boundary treated in~\cite{BCR13}.
Indeed, for $\xi^-=0$ and $\xi^+\neq 0$, the eigenvalues and the Bethe equations become independent of
$\xi^+$ as expected.
For the eigenvectors, one must notice that $\lim\limits_{\xi^-\rightarrow
0}\frac{\rho}{\xi^-}=-\frac{\xi^+}{2}$.
Then, using the Bethe equations, we recover the transmission coef\/f\/icients used in~\cite{BCR13} at least
for the small chains.

We conclude this section by a~technical complication due to the fact that the number of Bethe roots
$\lambda_j$ is equal to the length of the chain.
Indeed, let us notice that the vectors appearing in~\eqref{eq:f2} with total spin $N/2-m$ are
\begin{gather*}
\big\{\mathscr{B}(\lambda_{i_1})\mathscr{B}(\lambda_{i_2})\cdots\mathscr{B}(\lambda_{i_m})|\Omega\rangle
\big|
1\leq i_1<i_2<\dots<i_m\leq N\big\}
\qquad
\text{for}
\quad
m=0,1,\dots,N.
\end{gather*}
By simple enumeration, one shows that they span\footnote{These vectors are independent for generic
$\lambda_j$ i.e.\ when the $\lambda$ do not satisfy the Bethe equations.} a~vector space of dimension
$\genfrac{(}{)}{0pt}{}{N}{m}$ which is exactly the dimension of the sector with total spin $N/2-m$.
In the algebraic Bethe ansatz procedure, there are also vectors of the following type which appear
\begin{gather*}
\mathscr{B}(u)\mathscr{B}(\lambda_{j_2})\cdots\mathscr{B}(\lambda_{j_{m}})|\Omega\rangle
\qquad
\text{for}
\quad
1\leq j_2<\dots<j_m\leq N.
\end{gather*}
These vectors are also in the sector with total spin $N/2-m$.
Therefore, the latter vector is not independent of the former vectors and we have, for $1\leq j_2< \dots
<j_m\leq N$,
\begin{gather}
\mathscr{B}(u)\mathscr{B}(\lambda_{j_2})\cdots\mathscr{B}(\lambda_{j_{m}}
)|\Omega\rangle
\nonumber
\\
\qquad
=\sum_{1\leq i_1<\dots<i_m\leq N}V^{i_1,\dots,i_m}_{j_2,\dots,j_m}
(u,\lambda_1,\dots,\lambda_N)\mathscr{B}(\lambda_{i_1})\cdots\mathscr{B}(\lambda_{i_m})|\Omega\rangle,
\label{eq:extra}
\end{gather}
where $V$ are functions to be determined.
Let us illustrate this point by taking the particular case $m=N$.
In this case, one knows that
\begin{gather*}
\mathscr{B}(x_1)\mathscr{B}(x_2)\dots\mathscr{B}(x_N)|\Omega\rangle=Z^N(x_1,\dots,x_N)|\overline{\Omega}\rangle
\end{gather*}
where $Z^N$ is the partition function computed in~\cite{Tsu} and $|\overline{\Omega}\rangle$ is the vector
with all spins down.
Therefore, for example, we have
\begin{gather*}
V^{1,2,\dots,N}_{2,\dots,N}(u,\lambda_1,\dots,\lambda_N)=\frac{Z^N(u,\lambda_2,\dots,\lambda_N)}
{Z^N(\lambda_1,\lambda_2,\dots,\lambda_N)}.
\end{gather*}
Unfortunately, we do not know the functions $V$ for any $m$.
This point must be solved if one wants a~general proof of our proposal.

\section{Cases for chain with small length}\label{sec:smallL}

In this section, we give some details of the proof that~\eqref{eq:vpBE} is valid for $N=1,2,3$.

\subsection[Case $N=1$]{Case $\boldsymbol{N=1}$}

For $N=1$, the Bethe vector~\eqref{eq:eigen}, using relations~\eqref{eq:act}, becomes
\begin{gather*}
|\Phi^1(\lambda_1)\rangle=\overline{\mathscr{B}}(\lambda_1)|\Omega\rangle=\left(\mathscr{B}
(\lambda_1)+\frac{\rho}{\xi^-}\left(\frac{2\lambda_1}{2\lambda_1+1}\mathscr{A}(\lambda_1)-\mathscr{D}
(\lambda_1)\right)-\left(\frac{\rho}{\xi^-}\right)^2\mathscr{C}(\lambda_1)\right)|\Omega\rangle
\\
\phantom{|\Phi^1(\lambda_1)\rangle}
=\mathscr{B}(\lambda_1)|\Omega\rangle+W(\lambda_1)|\Omega\rangle,
\end{gather*}
with
\begin{gather*}
W(\lambda)=\frac{\rho}{\xi^-}\left(\frac{2\lambda}{2\lambda+1}\Lambda_1(\lambda)-\Lambda_2(\lambda)\right).
\end{gather*}
Now, let us write the action of the transfer matrix~\eqref{trane} on this Bethe vector by using the
exchange relations of Appendix~\ref{App:cr}:
\begin{gather*}
t(u)|\Phi^1(\lambda_1)\rangle=\Lambda^{\diag}(u)\mathscr{B}(\lambda_1)|\Omega\rangle+U^{\diag}
(u,\lambda_1)\mathscr{B}(u)|\Omega\rangle
+\xi^-(u+1)W(\lambda_1)\mathscr{B}(u)|\Omega\rangle
\\
\phantom{t(u)|\Phi^1(\lambda_1)\rangle=}
{}+\left(\big(\bar\alpha^{\diag}(u)\Lambda_1(u)+\bar\delta^{\diag}
(u)\Lambda_2(u)\big)W(\lambda_1)+\xi^+(u+1)G(u,\lambda_1)\right)|\Omega\rangle
\end{gather*}
where
\begin{gather*}
\bar\alpha^{\diag}(u)=\lim_{\rho\rightarrow 0}\alpha(u),
\qquad
\bar\delta^{\diag}(u)=\lim_{\rho\rightarrow 0}\delta(u),
\\
U^{\diag}(u,\lambda)=\left(\bar\alpha^{\diag}(u)g(u,\lambda)+\bar\delta^{\diag}
(u)n(u,\lambda)\right)\Lambda_1(\lambda)
\\
\phantom{U^{\diag}(u,\lambda)=}
{}+\left(\bar\alpha^{\diag}(u)w(u,\lambda)+\bar\delta^{\diag}(u)k(u,\lambda)\right)\Lambda_2(\lambda)
\end{gather*}
and
\begin{gather*}
G(u,\lambda)=\Lambda_1(u)\left(\big(m(u,\lambda)+l(u,\lambda)\big)\Lambda_1(\lambda)+p(u,\lambda)\Lambda_2(\lambda)\right)
\\
\phantom{G(u,\lambda)=}
{}+\Lambda_2(u)\left(\big(q(u,\lambda)+y(u,\lambda)\big)\Lambda_1(\lambda)+z(u,\lambda)\Lambda_2(\lambda)\right).
\end{gather*}

The function $U^{\diag}$ is exactly the term called unwanted term for the diagonal case and it is well-known
that it is proportional to the diagonal Bethe equation.
It reads here
\begin{gather*}
U^{\diag}(u,\lambda_1)=F(u,\lambda_1)\BE ^{\diag}_1.
\end{gather*}
As explained previously, the vectors $\mathscr{B}(u)|\Omega\rangle$ and
$\mathscr{B}(\lambda_1)|\Omega\rangle$ are not anymore independent for $N=1$ and we get
\begin{gather*}
\mathscr{B}(x)|\Omega\rangle=Z^1(x)|\overline\Omega\rangle
\qquad
\text{with}
\qquad
Z^1(x)=2x(p-\theta_1).
\end{gather*}
Then, we can compute, using~\eqref{eq:fr},
\begin{gather*}
(t(u)-\Lambda(u))|\Phi^1(\lambda_1)\rangle-F(u,\lambda_1)\BE_1|\Phi^1(u)\rangle
\\
\qquad
=\Big[-\rho\Lambda^{\gen}(u)Z^1(\lambda_1)+\left(-\rho F(u,\lambda_1)\BE_1^{\gen}
+\xi^-(u+1)W(\lambda_1)\right)Z^1(u)\Big]|\overline\Omega\rangle
\\
\qquad
\phantom{=}
{}+\Big[\big(\bar\alpha^{\diag}(u)\Lambda_1(u)+\bar\delta^{\diag}
(u)\Lambda_2(u)-\Lambda(u)\big)W(\lambda_1)
\\
\qquad
\phantom{=}
{}+\xi^+(u+1)G(u,\lambda_1)-F(u,\lambda_1)\BE_1W(u)\Big]|\Omega\rangle.
\end{gather*}
Finally, using the explicit expressions of the functions, one can show that the terms in front of~$|\Omega\rangle$ and~$|\overline \Omega\rangle$ vanish.
This proves~\eqref{eq:vpBE} and validates our proposal for $N=1$.

\subsection[Case $N=2$]{Case $\boldsymbol{N=2}$}

For $N=2$, the Bethe vector~\eqref{eq:eigen} becomes
\begin{gather*}
|\Phi^2(\lambda_1,\lambda_2)\rangle=\mathscr{B}(\lambda_1)\mathscr{B}
(\lambda_2)|\Omega\rangle+W_1(\lambda_1,\lambda_2)\mathscr{B}
(\lambda_1)|\Omega\rangle+W_2(\lambda_1,\lambda_2)\mathscr{B}(\lambda_2)|\Omega\rangle
\\
\phantom{|\Phi^2(\lambda_1,\lambda_2)\rangle=}
{}+W_{\varnothing}(\lambda_1,\lambda_2)|\Omega\rangle,
\end{gather*}
where the explicit expressions of the functions $W_1, W_2$ and $W_{\varnothing}$ are
\begin{gather*}
W_1(\lambda_1,\lambda_2)=\frac{\rho}{\xi^-}\left(\frac{2\lambda_2}{2\lambda_2+1}
\Lambda_1(\lambda_2)f(\lambda_2,\lambda_1)-\Lambda_2(\lambda_2)h(\lambda_2,\lambda_1)\right),
\\
W_2(\lambda_1,\lambda_2)=W_1(\lambda_2,\lambda_1)
\end{gather*}
and
\begin{gather*}
W_{\varnothing}(\lambda_1,\lambda_2)=\left(\frac{\rho}{\xi^-}\right)^2\Bigg(\frac{\lambda_1+\lambda_2+2}
{\lambda_1+\lambda_2+1}\Lambda_2(\lambda_1)\Lambda_2(\lambda_2)-\frac{2\lambda_2}{2\lambda_2+1}
\frac{\lambda_1-\lambda_2+1}{\lambda_1-\lambda_2}\Lambda_2(\lambda_1)\Lambda_1(\lambda_2)
\\
\qquad{}
-\frac{2\lambda_1}{2\lambda_1+1}\frac{\lambda_2-\lambda_1+1}{\lambda_2-\lambda_1}
\Lambda_1(\lambda_1)\Lambda_2(\lambda_2)+\frac{2\lambda_1}{2\lambda_1+1}\frac{2\lambda_2}{2\lambda_2+1}
\frac{\lambda_1+\lambda_2}{\lambda_1+\lambda_2+1}\Lambda_1(\lambda_1)\Lambda_1(\lambda_2)\Bigg).
\end{gather*}

The computation for $N=2$ follows the same lines as the case $N=1$ and is very similar to the usual ones.
The crucial point is that the three vectors $\mathscr{B}(\lambda_1)|\Omega\rangle$,
$\mathscr{B}(\lambda_2)|\Omega\rangle$ and $\mathscr{B}(u)|\Omega\rangle$ are not independent.
We can show that
\begin{gather*}
\mathscr{B}(u)|\Omega\rangle=V^1(u,\lambda_1,\lambda_2)\mathscr{B}
(\lambda_1)|\Omega\rangle+V^2(u,\lambda_1,\lambda_2)\mathscr{B}(\lambda_2)|\Omega\rangle,
\end{gather*}
where
\begin{gather*}
V^1(u,\lambda_1,\lambda_2)=\frac{u(u-\lambda_2)(u+\lambda_2+1)}
{\lambda_1(\lambda_1-\lambda_2)(\lambda_1+\lambda_2+1)}
\end{gather*}
and $V^2(u,\lambda_1,\lambda_2)=V^1(u,\lambda_2,\lambda_1)$.
Using this relation, we prove that our proposal is valid also for $N=2$.

\subsection[Case $N=3$]{Case $\boldsymbol{N=3}$}

For $N=3$, the proof follows the same line as previously.
The new feature lies on the relations between the vectors.
Therefore, one gets
\begin{gather*}
\mathscr{B}(u)|\Omega\rangle
=V^1(u,\lambda_1,\lambda_2,\lambda_3)\mathscr{B}(\lambda_1)|\Omega\rangle
+V^2(u,\lambda_1,\lambda_2,\lambda_3)\mathscr{B}(\lambda_2)|\Omega\rangle
\\
\phantom{\mathscr{B}(u)|\Omega\rangle=}
{}+V^3(u,\lambda_1,\lambda_2,\lambda_3)\mathscr{B}(\lambda_3)|\Omega\rangle,
\end{gather*}
where
\begin{gather*}
V^1(u,\lambda_1,\lambda_2,\lambda_3)=\frac{u(u-\lambda_2)(u+\lambda_2+1)(u-\lambda_3)(u+\lambda_3+1)}
{\lambda_1(\lambda_1-\lambda_2)(\lambda_1+\lambda_2+1)(\lambda_1-\lambda_3)(\lambda_1+\lambda_3+1)}
\end{gather*}
and $V^2(u,\lambda_1,\lambda_2,\lambda_3)=V^1(u,\lambda_2,\lambda_1,\lambda_3)$,
$V^3(u,\lambda_1,\lambda_2,\lambda_3)=V^1(u,\lambda_3,\lambda_1,\lambda_2)$.
One gets also the following relations
\begin{gather*}
\mathscr{B}(u)\mathscr{B}(\lambda_1)|\Omega\rangle=V^{12}_1(u,\lambda_1,\lambda_2,\lambda_3)\mathscr{B}
(\lambda_1)\mathscr{B}(\lambda_2)|\Omega\rangle+V^{13}_{1}(u,\lambda_1,\lambda_2,\lambda_3)\mathscr{B}
(\lambda_1)\mathscr{B}(\lambda_3)|\Omega\rangle
\\
\phantom{\mathscr{B}(u)\mathscr{B}(\lambda_1)|\Omega\rangle=}
{}+V^{23}_1(u,\lambda_1,\lambda_2,\lambda_3)\mathscr{B}(\lambda_2)\mathscr{B}(\lambda_3)|\Omega\rangle,
\end{gather*}
where the functions $V$ in this formula is too long to be displayed here.
We used formal mathematical software to compute it.
Unfortunately, we do not f\/ind yet a~closed formula for them.
Using these relations, we prove that our proposal is valid also for $N=3$.

\section{Conclusion}\label{sec5}

In this paper, we construct the Bethe vectors, thanks to the algebraic Bethe ansatz, for the Heisenberg
spin chain with generic boundaries.

A future task is to prove our conjecture.
For that, it should be necessary to obtain the explicit formulas for the functions $V$ present
in~\eqref{eq:extra}.
Then, we believe that the computation of the correlation functions would be possible following the same
lines as the diagonal case.

The study of the thermodynamic limit ($N\rightarrow\infty$) would be also interesting to compute the ground
state energy, the f\/irst excited states and the dif\/fusion matrices between them or on the boundary.
For that, it is necessary to identify the solution of the Bethe equations~\eqref{eq:BEK} corresponding to
the ground state and to get the integral equations for the density of the Bethe roots.
Then, one should compute the thermodynamic limit for the correlation functions.

We believe also that our proposal may be applied to get the eigenvectors for various other problems: $(i)$~the twisted closed XXX spin chain has been studied by one of the author~\cite{Bel}; $(ii)$~the open XXX spin
chain with spin $s$: the dif\/f\/iculty here would be again to compute the equivalent of the functions $V$;
$(iii)$~the XXZ spin chains and to study their relation with the asymmetric simple exclusion process.
We expect that our construction can be used directly to study the case with $K^-$ diagonal and $K^+$
general as the method used to study the triangular boundaries for XXX model~\cite{BCR13} has been
generalized to XXZ model~\cite{pimenta}.
However, in the XXZ case, the $R$-matrix is not anymore invariant and one cannot diagonalize the $K^-$-matrix.
Therefore to deal with both general boundaries one must generalize our approach but we believe that our
proposal is a~f\/irst step to get the eigenvectors; $(iv)$~the higher rank spin chains; $(v)$~other type of
integrable models with non diagonal twist or boundary.

\appendix

\section{Exchange relations
\label{App:cr}
}

Using ref\/lection equation~\eqref{eq:re}, we can f\/ind the exchange relations between the operators
$\mathscr{A}$,~$\mathscr{B}$,~$\mathscr{C}$ and $\mathscr{D}$.
For our calculations, we only need the following ones
\begin{gather*}
[\mathscr{B}(u),\mathscr{B}(v)]=0,
\\
\mathscr{A}(u)\mathscr{B}(v)=f(u,v)\mathscr{B}(v)\mathscr{A}(u)+g(u,v)\mathscr{B}(u)\mathscr{A}
(v)+w(u,v)\mathscr{B}(u)\mathscr{D}(v),
\\
\mathscr{D}(u)\mathscr{B}(v)=h(u,v)\mathscr{B}(v)\mathscr{D}(u)+k(u,v)\mathscr{B}(u)\mathscr{D}
(v)+n(u,v)\mathscr{B}(u)\mathscr{A}(v),
\\
\null[\mathscr{C}(u),\mathscr{B}(v)]=m(u,v)\mathscr{A}(v)\mathscr{A}(u)+l(u,v)\mathscr{A}(u)\mathscr{A}
(v)+q(u,v)\mathscr{A}(v)\mathscr{D}(u)
\\
\phantom{\null[\mathscr{C}(u),\mathscr{B}(v)]=}
{}+p(u,v)\mathscr{A}(u)\mathscr{D}(v)+y(u,v)\mathscr{D}(u)\mathscr{A}(v)+z(u,v)\mathscr{D}(u)\mathscr{D}(v),
\end{gather*}
with
\begin{alignat}{3}
& f(u,v)= \frac{(u-v-1)(u+v)}{(u-v)(u+v+1)},
\qquad &&
 h(u,v)= \frac{(u-v+1)(u+v+2)}{(u-v)(u+v+1)},&
\nonumber
\\
& w(u,v)= \frac{-1}{(u+v+1)},
\qquad &&
g(u,v)= \frac{2v}{(2v+1)(u-v)}, &
\nonumber
\\
& k(u,v)=\frac{-2(u+1)}{(u-v)(2u+1)},
\qquad &&
n(u,v)= \frac{4v(u+1)}{(u+v+1)(2v+1)(2u+1)},&
\label{eq:fg}
\end{alignat}
and
\begin{alignat*}{3}
& m(u,v)=\frac{2 u(u-v+1)}{(2u+1)(u+v+1)(u-v)},
\qquad &&
l(u,v)=-\frac{2 u}{(2u+1)(2v+1)(u-v)}, &
\\
& q(u,v)=\frac{(u+v)}{(u+v+1)(u-v)},
\qquad &&
p(u,v)=-\frac{2 u}{(2u+1)(u-v)}, &
\\
& y(u,v)=-\frac{1}{(u+v+1)(2v+1)},
\qquad &&
z(u,v)=-\frac{1}{u+v+1}. &
\end{alignat*}

\subsection*{Acknowledgements}

We would like to thank P.~Baseilhac, V.~Caudrelier, R.A.~Pimenta and E.~Ragoucy for discussions.

\pdfbookmark[1]{References}{ref}
\LastPageEnding

\end{document}